\documentclass[final,endfloats,superscriptaddress,aps,pre,a4]{revtex4}
\usepackage{latexsym}
\usepackage{times}
\usepackage{graphicx}
\usepackage{amsmath}
\usepackage{amsxtra}
\usepackage{amstext}
\usepackage{amssymb}
\usepackage{latexsym}
\usepackage{dsfont}

\begin{document}

\author{D.~Fusco}
\affiliation{Universit\`a degli Studi di Milano, Dip.
    Fisica, Via Celoria 16, 20133 Milano, Italy }
\author{B.~Bassetti}
\affiliation{Universit\`a degli Studi di Milano, Dip.
    Fisica, Via Celoria 16, 20133 Milano, Italy }
\affiliation{I.N.F.N., Milano, Italy}
\email[e-mail address: ]{bassetti@mi.infn.it}
\author{P.~Jona}
\affiliation{Politecnico di Milano, Dip. Fisica, Pza Leonardo Da Vinci
  32, 20133 Milano, Italy}
\author{M.~Cosentino Lagomarsino}
\affiliation{UMR 168 / Institut Curie, 26 rue d'Ulm 75005 Paris, France}
\affiliation{Universit\`a degli Studi di Milano, Dip.
    Fisica, Via Celoria 16, 20133 Milano, Italy }
\email[ e-mail address: ]{mcl@curie.fr}
%

\title{DIA-MCIS. An Importance Sampling Network Randomizer for Network Motif
  Discovery and Other Topological Observables in Transcription Networks. }

\maketitle

\section{Abstract}

\textbf{Motivation:} Transcription networks, and other directed
networks can be characterized by some topological observables such
as for example subgraph occurrence (network motifs).  In order to
perform such kind of analysis, it is necessary to be able to
generate suitable randomized network ensembles. Typically, one
considers null networks with the same degree sequences of the
original ones. The commonly used algorithms sometimes have long
convergence times, and sampling problems.  We present here an
alternative, based on a variant of the importance sampling
Montecarlo developed by Chen \emph{et al.}~\cite{diaconis}.

\bigskip

\textbf{Availability:} softwares are available at
http://wwwteor.mi.infn.it/~bassetti/downloads.html

\bigskip

\textbf{Contact:} marco.cosentino-lagomarsino@curie.fr,
diana.fusco@studenti.unimi.it

\bigskip

\textbf{Supplementary Information:} supplementary notes are
available at http://wwwteor.mi.infn.it/~bassetti/downloads.html

\section{Introduction}

Gene regulatory networks are graphs that represent interactions
between genes or proteins.  For example, in a transcription network
the nodes are genes or operons, identified with their protein
products, and the edges represent their transcriptional regulatory
regions along DNA~\cite{babu}. The simplest possible approach to
study them is to consider their topology.  The main biological
question that underlies these studies asks to establish to what
extent the empirical biological topology deviates from a ``typical
case'' statistics.
In order to do that, one generates so called ``randomized
counterparts'' of the original data set as a null model. That is, an
ensemble of random networks which conserve some topological
observables of the original, such as the degree sequences, i.e. the
number of outgoing and incoming links for each node.
This approach has a wider application for networks of different
kinds~\cite{MIK04,alon}. A directed network can be conveniently
represented as a zero-one adjacency matrix where element $a_{(i,j)}$
is 1 if node $j$ has a directed link pointing to node $i$ (Fig.1A).
The null ensemble of degree-conserving graphs translates into a set
of matrices having the same row and column sums of the original
matrix. Some algorithms to generate this uniformly distributed
ensemble are commonly used~\cite{diaconis,alon1}.  In particular,
one Markov Chain Montecarlo (MCMC) algorithm is based on swapping
edges at random~\cite{MR95}. This generates an ergodic dynamics,
with, however, large relaxation times to a uniform distribution.
Another type of algorithm is the so called ``stub-pairing'' or
Molloy-Reed algorithm~\cite{MR95,alon1}, that consists in randomly
linking ``stubs'' made of nodes with required in- and outdegrees, in
order to build a randomized instance~\cite{RJB96,MS05}. While
useful, this technique may fall in metastable states, where no stubs
can be connected.  The algorithm developed by Chen \emph{et
al.}~\cite{diaconis} is more efficient than the MCMC one
\cite{diaconis} and does not run the risk of falling in metastable
states.  It is based on an application of importance sampling
Montecarlo. It generates matrices with an almost uniform
probability, and subsequently adjusts the sample, assigning to every
element a certain weight. Finally, it is able to estimate the size
of the sampled ensemble.

Here, we present an implementation of this algorithm that works specifically
on transcription networks, with two variants. The first variant is designed to
improve the speed of the algorithm. The second variant enables to deal with
ensembles of structured matrices, in particular with structured diagonal, as
it is often done in transcription networks when dealing with
self-regulations~\cite{alon}.

\section{Algorithm}

As the goal is the uniform distribution of the sample, the
importance sampling weight for every element is $1/P(T)$, where
$P(T)$ is the matrix probability.
The algorithm is illustrated in Fig.1A. The matrix is generated by
filling column after column. Suppose, for example, the first column
has already been generated and the second one (in pink in Fig.1A)
must be extracted. One has to consider the row sums having
subtracted the first column. At this point, one can compute a
"constraint" inside the column in order to allow the algorithm not
to fall in metastable states (Fig.1A and~\cite{diaconis}).
Subsequently, the constraint-free positions are filled with a
probability that can be  computed exactly~\cite{liu}. In order to
perform this operation, the row sums need to be ordered by rank.
When all the columns are filled, the total probability of having a
certain matrix is the product of all the column probabilities, which
can be computed knowing the constraints of each column~\cite{liu}.
This number allows to weigh correctly the matrix sample.

We introduced the following two variants.

\bigskip

\textbf{Large matrices with compact indegree} Transcription networks
typically have several hundreds of nodes.  The computational cost
for generating a column is of order $\mathcal{O}(M^2c^2)$ where $M$
is the length of a column and $c$ the number of 1s contained in that
column~\cite{diaconis}. This is due to the fact that every time that
a position must be selected, the algorithm has to evaluate the
probability of success for every position inside the
column~\cite{vigoda}.

We have demonstrated that the probability of success in a given
position can be well approximated using the corresponding row-sum if
the in-degree distribution is sufficiently limited in range. This
last feature is typical of transcription networks. Consequently, as
the probability of having a certain zero-one sequence does not
depend on the order of extraction, it can be evaluated only once for
every column, or, better, for each constraint. The computational
cost for generating a column is then reduced to order
$\mathcal{O}(Mc)$.

\bigskip

\textbf{Structured diagonal} Self-regulatory interactions are often
considered to have a particular status \cite{alon}. They are
represented in the matrix by 1 on the diagonal~\cite{alon}. In order
to constrain the diagonal, one has to modify the way the algorithm
calculates the constraints inside the columns, accounting for the
fact that some positions are not available for the extraction.

\bigskip

\begin{figure*}[!ht]
\centering
  \includegraphics[width=0.9\textwidth]{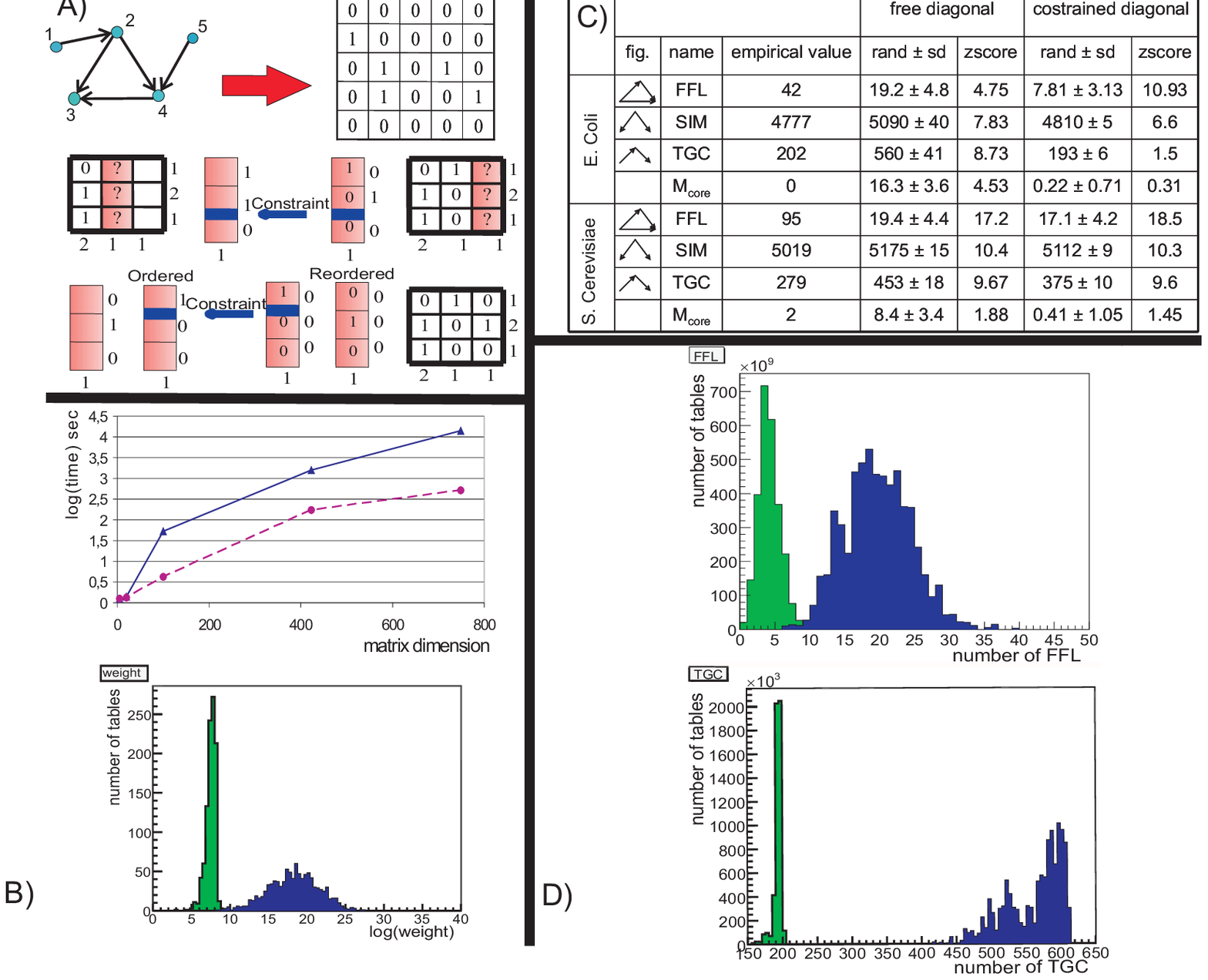}
  \caption{A) Description of the algorithm. Top: the graph is translated
    into an adjacent matrix, which is filled column by column. Middle/Bottom:
     the procedure for generating a matrix is showed. The second column
    (in pink) must be generated. In order to perform this operation, the updated row
    sums and the constraint are calculated. The free positions are
    extracted and the algorithm considers the next (and last) column. Bottom: the column is then reordered by the residual row sums.
     One constraint is found and the last column filled. The matrix is now complete.
    B) Performance and uniformity of the sample. Top: comparison of the
    computational cost of our variant with the algorithm of Chen \emph{et
    al.}.
    We plotted (in logarithmic scale) the time to generate 100 randomizations
    of networks of different size (a random subgraph of the
    \textit{E. Coli} network, the entire \textit{E. Coli} network,
    the \textit{B.Subtilis} network\cite{DBTBS}).
    The solid line (blue) represents the algorithm of Chen \emph{et al.}, the
    dashed line (purple) represents our modified algorithm. Bottom: weight distribution of the two
    algorithms for a sample of 1000 randomizations of the \textit{E.  Coli}
    network.  The narrower the distribution of weights, the better the
    algorithm approximates a uniform sample. Our modified algorithm (light
    green) tends to generate matrices that have a higher probability of extraction
     than the other one (dark blue).  This explains why its
    distribution is narrower.
    C) Table summarizing the results for three
    triangular motifs and the feedback (measured by $M_{core}$\cite{gammachi} of the \emph{E.~coli} and \emph{S.
      Cerevisiae} transcription networks. Note that the \textit{E. Coli} dataset has no feedback.
      D) Comparison of subgraph (FFL, TGC) distributions in \textit{E.~coli}
    randomizations for structured (light green) and unstructured (dark blue)
    diagonals.  These distributions are systematically shifted with respect to
    each other.}
\end{figure*}

\section{Implementation and results}

\textbf{Triangular network motifs} As an example of application we
have studied the occurrence of three triangular subgraphs (Fig.1C
and 1D). The FFL (Feed Forward Loop), SIM (triangular Single Input
Module) and TGC (Three Gene Chain), for the transcription networks
of \emph{E.~coli}~\cite{shenn} and \emph{S. Cerevisiae}~\cite{data}
verifying the results that can be found in the
literature~\cite{alon,MIK04}.

In all cases, we find a quantitative difference between the subgraph
distributions in the randomized ensembles with or without structured
diagonal (Fig.1C and 1D). In some instances, such as the
biologically relevant FFL~\cite{alon}, this does not affect the
decision of whether that subgraph is a motif. In other cases one can
also find qualitative changes. This difference is more visible in
\textit{E. Coli} as sixty percent of its nodes are autoregulated,
and less in \textit{S. Cerevisiae} with only ten percent of
autoregulations.

\bigskip

\textbf{Feedback} We also evaluated (Fig.1C) the feedback in the
graph, using a simple decimation algorithm that removes the input-
and output- treelike components~\cite{gammachi}. With this
algorithm, the feedback is measured by the size $M_{core}$ of the
decimated graph. We have ignored autoregolations. As expected, the
sample with structured diagonal is shifted towards smaller amounts
of feedback. This can be explained considering the lower amount of
available links to rearrange if the selfregulators are fixed.

\bigskip

\section{Conclusions}

In conclusion, we have implemented a Montecarlo importance sampling
algorithm to randomize directed graphs conserving the degree
sequence, and evaluate topological observables. The algorithm
follows the design principles of Chen \emph{et al.}, and is designed
to be more efficient without loss of uniformity on graphs with
compact indegree such as the known transcription networks.
Furthermore, we added a variant that works with constrained
diagonal, as is usually done in motif discovery~\cite{alon}.  We
implemented the code in a simple three-node motif and feedback
finder, that reproduces the results known in the literature.  The
version of the running code (in $C^{++}$) used for our analysis is
publicly available at
http://wwwteor.mi.infn.it/~bassetti/downloads.html , and can be
inserted in more general motif finding tools.

\section*{Acknowledgement}

The authors would like to thank F. Bassetti, S. Holmes and P.
Diaconis for helpful discussion.

\section{Additional Notes on the Importance Sampling Randomizer for
  Transcription Networks}

\subsection{Introduction}

The purpose of these notes is to introduce and describe two
modifications of the importance sampling randomization algorithm for
directed graphs introduced in~\cite{diaconis}.
The sample of randomized graphs to be generated has to be uniform in
the set of graphs having the same degree sequences as the original
one, i.e. conserving the number of incoming and outgoing edges for
each node~\cite{alon1}.
These modifications are produced keeping in mind two important
features of transcriptional regulation networks.  The first is that
these graphs have compact indegree. For example in the case of the
Shen-Orr~\cite{alon1} data-set for the~\emph{E.~coli} transcription
network, a graph with about 400 nodes and 600 edges, the maximum
indegree is of order 10, while the maximum outdegree has order 100.
The second feature is that networks may have an abundance of
self-interactions (this is the case for example in \emph{E.~coli}).
For this reason, one may wish to consider randomizations that
conserve the number of self-interactions, i.e. having structured
diagonal in the adjacency matrix (see below).

\subsection{Summary of the Procedure}

A directed graph can be represented by an adjacency matrix $G$ where
the element indexed by $(i,j)$ is 1 if gene $j$ influences gene $i$,
and 0 otherwise. Row sums of the matrix represent the number of
nodes receiving edges from each node (outdegree), column sums
represent the number of nodes sending edges to each node (indegree).
Consequently, generating randomized networks with fixed in- and
outdegree is equivalent to generating randomized matrices with
constant row- and column sums.

The algorithm of~\textit{Chen et al.}\cite{diaconis} has this scope,
and achieves it using the Montecarlo importance sampling method:
every matrix is generated column by column and is then weighted
inside the sample with a certain analytically calculated weight.
This weight consists in the inverse of the probability that the
matrix is generated by the algorithm. The calculation of the matrix
probability is a crucial point. It is performed using the
conditional Poisson distribution~\cite{liu}.
This distribution allows to compute the probability of having a 0-1
sequence of length $N$ with the constraint of having $n$ nonzero
entries. A key role is played by the function
\begin{equation}
\mathcal{R}(k,C)= \left\{
\begin{array}{ll}
1 & \mbox{if } k=0\\
\sum_{B\subset C,|B|=k}(\prod_{i\in B}w_i) & \mbox{if } 0 <k \leq |C|\\
0 & \mbox{if } k > |C|
\end{array}
\right.
\end{equation}
where $C$ is the set of the possible positions in the sequence (in
this case $C=\{1,...,N\}$ and $w_i$ is the weight assigned to
position $i$. When a column is generated, this weight is $r_i/N$,
where $r_i$ is the $i$th row sum.

Suppose now that the positions where 1 are put are extracted one by
one and that $A_k$ is the set that contains the positions chosen
after the $k$th extraction. At the beginning $A_0=\emptyset$. Then
at the $k$th step the position $j\in A_{k-1}$ will be extracted with
probability
\begin{equation}\label{eqn:estrazione}
P(j,A_{k-1}^c)=\frac{w_j \mathcal{R}(n-k,A_{k-1}^c\backslash
j)}{(n-k+1)\mathcal{R}(n-k+1,A_{k-1}^c)} \ ,
\end{equation}
where $w_j=\frac{p_j}{1-p_j}$ is the weight assigned to position
$j$.

\subsection{Large Matrices}

The first problem we had to face was due to the dimensions of our
matrices. The networks we considered typically had about 500 nodes,
consequently the associated matrix is $500\times 500$.  With these
number, the algorithm of Chen \emph{et. al.} is too slow to generate
a significant sample in reasonable time. Now, most of the computing
time is required by the calculation of $\mathcal{R}(c,A)$.

To avoid this problem, we use the following method.  Suppose that
the $l$th column is being generated, and it has to contain $c$
edges, or units. Then for every row with $r_j^{(l)}\neq 0$, at least
the numerator (it depends on $j$) of Eq.~\ref{eqn:estrazione} must
be calculated. The denominator is a common factor to all the rows,
and is not important at this step. A similar calculation has to be
performed for every placement from 1 to $c$.  The process for
calculating $\mathcal{R}(c-k,A_{k-1}^c)$ has a computational cost of
order $\mathcal{O}((c-k)*|A_{k-1}^c|)$~\cite{liu}. This calculation
must be repeated for every available position $j$ that is of order
$|A_{k-1}^c|$.  To avoid repeating the process for all the $c$
extractions, we approximate $A_{k-1}^c$ with the number of rows $M$
(typically about 500) of the matrix, for every $k$, then the cost
for generating a column becomes of order $\mathcal{O}(M^2c^2)$. In
other words, approximating the probability of selection of a certain
position with its row sum, the algorithm should calculate the
function $\mathcal{R}$ only once for every column, reducing
considerably the computational cost.
We will now argue that this approximation is acceptable for graphs
with ``small'' indegree.

The probability of selecting a string
$(A_{l,1}=a_1,...,A_{l,M}=a_M)$ with prescribed sum does not depend
on the extraction order. It simply writes
\begin{equation}\label{eqn:colonna}
P(A_{l,1}=a_1,...,A{l,M}=a_M|\sum_{i=1}^M a_i=c)=\frac{\prod_{i=1}^M
w_i^{a_i}}{\mathcal{R}(c,C)}
\end{equation}
where $C$ is the set that represent the whole column.
This means that for evaluating this probability one does not have to
keep into account the whole process of extraction.
However, the problem of making a good extraction still persists. In
fact, even if the calculation of the sequence probability is
correct, nothing assures that this sequence has been extracted with
the conditional Poisson distribution.  First of all note that the
statistical meaning of $\mathcal{R}(n,C)$ is:
\begin{equation}
\mathcal{R}(n,C)=\frac{P(S_C=n)}{P(S_C=0)} \ ,
\end{equation}
where the random variable $S_C=\sum_{i=1}^M a_i$.
Consequently, if we compare the probabilities of extracting the
position $i$ and the position $j$ at the $k$th step, they can be
written as

\begin{equation}
\left\{
\begin{array}{lll}
p(i)&\propto&r_i P(S_i=c-k)\\
p(j)&\propto&r_j P(S_j=c-k)  \  , \\
\end{array}
\right.
\end{equation}

where $S_i$ stands for the sum of the elements of $A_{k-i}^c
\setminus i$ and $S_j$ stands for the sum of the elements of
$A_{k-i}^c \setminus j$.

Now, note that
\begin{equation}
P(S_i=c-k)=\sum_{B\subset C\backslash i,|B|=c-k}\prod_{x\in
B}p_x\prod_{y\in B^c}(1-p_y) \ \ .
\end{equation}

Among all the sets $B$, there will be some that contain $j$.
Equivalently, for $S_j$, there will be some sets $B$ containing $i$
and some not containing it. As the sum runs over all the possible
subsets, we can write it as follows

\begin{equation}\label{eqn:differenza}
\left\{
\begin{array}{lll}
P(S_i=c-k)&=&p_j\sum_{B\subset C\backslash i,|B|=c-k,j\in
B}\prod_{x\in B,x\neq j}p_x\prod_{y\in
B^c}(1-p_y)+\\
&+&(1-p_j)\sum_{B\subset C\backslash i,|B|=c-k,j\not\in
B}\prod_{x\in B}p_x\prod_{y\in B^c,y\neq j}(1-p_y)\\
P(S_j=c-k)&=&p_i\sum_{B\subset C\backslash j,|B|=c-k,i\in
B}\prod_{x\in B,x\neq i}p_x\prod_{y\in
B^c}(1-p_y)+\\
&+&(1-p_i)\sum_{B\subset C\backslash j,|B|=c-k,i\not\in
B}\prod_{x\in B}p_x\prod_{y\in B^c,y\neq i}(1-p_y)\\
\end{array}
\right. \ .
\end{equation}

Note that the factor multiplying $p_j$ in the first equation is the
same as the factor multiplying $p_i$ in the second ($\mathfrak{A}$).
The same happens for $1-p_j$ and $1-p_i$($\mathfrak{B}$). Thus, we
can rewrite equation \ref{eqn:differenza} as

\begin{equation}
\left\{
\begin{array}{lll}
P(S_i=c-k)&=&p_j\mathfrak{A}+(1-p_j)\mathfrak{B}\\
P(S_j=c-k)&=&p_i\mathfrak{A}+(1-p_i)\mathfrak{B}\\
\end{array}
\right. \ .
\end{equation}

If we now consider the difference between the two equations

\begin{equation}\label{eqn:chiave}
P(S_i=c-k)-P(S_j=c-k)=(p_j-p_i)(\mathfrak{A}-\mathfrak{B}) \ ,
\end{equation}
we see that $|\mathfrak{A}-\mathfrak{B}|\leqslant 1$, as separately
$\mathfrak{A}\leqslant 1$ and $\mathfrak{B}\leqslant 1$. Now
$p_j-p_i=\frac{r_j-r_i}{N-l}$ where $r_i$ and $r_j$ are the updated
row sums (updated after the genration of the previous $l-1$
columns). Then it is easy to see that $|p_j-p_i|\leq
\frac{r_{max}-2}{r_{max}}$ where $r_{max}= max_i r_i$. This is due
to the fact the worst situation is when for example $r_i=1$ and
$r_j=N-l-1$. As $r_j \leq r_{max}$ the most approximated step is
when $N-l=r_{max}$. This explains why smaller values of $r_{max}$
lead to a better approximation. The probability of being in this
situation is proportional to the probability that, after $N-r_{max}$
generated columns, the column with the maximum row sum is empty
apart form one unit. In order to estimate it roughly, we consider
the rows as independent and approximate the row distribution with a
Bernoulli distribution with probability of success $r_{max}/N$, then
the probability of having a sequence of $N-r_{max}$ zeros is
estimated as:
$$
P=\frac{(N-r_{max})^{N-r_{max}}}{N^{N-r_{max}}} \ \ .
$$
This probability decreases if $r_{max}$ increases. For example, for
the \textit{E.~Coli} graph, it is equal to 0.00257, as $N=423$ and
$r_{max}=6$. This gives a rough estimate of the maximum error.

\subsection{Constrained Diagonal}

 Self-interactions (units on the diagonal
of $G$) have particular status in transcription
networks~\cite{alon}. For this reason, it is interesting to consider
randomized ensembles where the diagonal is constrained. The problem
is then how to make the diagonal inaccessible for the algorithm
column-filling steps, and in particular, how to calculate the
constrains inside the columns.

First, we note that the positions above the diagonal behave as in
the previous case. The problem restricts to are the positions below
the diagonal.  The algorithm to find the constraints inside every
column can be summarized as follows.
\begin{enumerate}
\item Order the position ranking them from the highest row-sum to the lowest.
\item If two or more positions have the same row-sums, the positions below the
  diagonal must be placed first.
\item Among the positions below the diagonal having the same row sums, a
  precise order must be followed. Suppose that after the previous ordering
  step row $i$ occupies position $P_i$. Then the rows with the lowest
  difference $|c_i-P_i|$ have the priority.
\item Let $\mathbf{P}^{-1}$ be the vector of positions before the ordering
  step, i.e. the row occupying now position $j$ is the row that occupied
  position $P_j^{-1}$ before reordering.  Considering the difference
  $\sum_{i=1}^k r_i-\sum_{i=1}^k c_i^{(2)*}$, one unit must be subtracted if
  $c_{P_k^{-1}}\geq k$ and if $P_k^{-1}$ is under the diagonal.
\item When $k$ becomes large enough so that for some $i$ $c_{P_i^{-1}}\leq k$,
  one unit for every $i$ must be summed. This must be done only if previously
  one unit had been subtracted for that positions.
\end{enumerate}

In this way the two vectors $\mathbf{K}$ and $\mathbf{v}$
identifying the constrains inside the columns will take into account
the inaccessibility of the diagonal.
Finally, while placing the units inside the columns, it must be kept
in mind that the positions of the diagonal are not accessible. This
must be considered also when assigning the weights to every row and
the probability of having a certain number of units before every
constraint.


\begin{thebibliography}{99}



\bibitem{diaconis} Y. Chen, P. Diaconis, S. P. Holmes, J. S. Liu.
  \emph{Sequential Monte Carlo Methods for Statistical Analysis of Tables}.
  Journal of the American Statistical Association, 100, 109-120, (March 2005).

\bibitem{liu} S. X. Chen, J. S. Liu.  \emph{Statistical applications of the
    Poisson-binomial and conditional Bernoulli distributions}.  Statistica
  Sinica 7, 875-892, (1997).

\bibitem{vigoda} I. Bekazova, A. Sinclair, D. Stefankovic, E.  Vigoda.
  \emph{Negative Examples for Sequential Importance Sampling of Binary
    Contingency Tables} in Y.  Azar, T. Erlebach(Eds): Algorithms-ESA 2006,
  14th Annual European Symposium, Zurich, Switzerland, September 11-13,2006,
  Proceedings. Lecture notes in Computer Science 4168 Springer, 136-147.

\bibitem{alon} N. Kashtan, S. Itzkovitz, R. Milo, U. Alon \emph{Efficient
    sampling algorithm for estimating subgraph concentrations and detecting
    network motifs} Bioinformatics 20(11), 1746-1758, (2004).

\bibitem{alon1} R.~Milo, N.~Kashtan, S.~Itzkovitz, M.~E.~J. Newman, U.~Alon,
  \emph{On the uniform generation of random graphs with prescribed degree
    sequences} {\it cond-mat/0312028\/} (2003).

\bibitem{MR95} M.~Molloy and B.~Reed, {\it Random Structures and
    Algorithms} \textbf{6}, 161-179 (1995).


\bibitem{RJB96}
A.~Rao, R.~Jana, S.~Bandyopadhyay, {\it Indian J. Stat.\/} {\bf 58(A)}, 225
  (1996).
\bibitem{MS05}
S.~Maslov, K.~Sneppen {\it Phys Biol\/} {\bf 2 (4)}, S94 (2005).

\bibitem{MSI+02}
R.~Milo, {\it et~al.\/}, {\it Science\/} {\bf 298}, 824 (2002).

\bibitem{MIK04}
R.~Milo, {\it et~al.\/}, {\it Science\/} {\bf 303}, 1538 (2004).

\bibitem{gammachi} Cosentino Lagomarsino, M., Bassetti B., Jona P.,
  {\it Lecture Notes in Bioinformatics}, Proceedings of
  the CMSB conference 2006. Springer-Verlag, 2006 (q-bio.MN/0606039).

\bibitem{babu} Madan Babu M., Luscombe N., Gerstein M., Aravind L Teichmann
  S.A.  {\it Structure and evolution of gene regulatory networks} {\it Curr.
    Opin. Struct. Biol.} {\bf 14}, 283-291 (2004).

\bibitem{data} Guelzim N, Bottani S, Bourgine P, Kepes F.Topological and
  causal structure of the yeast transcriptional regulatory network.  Nat
  Genet. 2002 May;31(1):60-3

 \bibitem{DBTBS} Makita Y, Nakao M, Ogasawara N, Nakai K.,
{ \bf DBTBS:} {\it database of transcriptional regulation in
Bacillus subtilis and its contribution to comparative genomics
Nucleic Acids Res.}, {\bf 32},D75-77 (2004)


\bibitem{shenn} Shen-Orr SS, Milo R, Mangan S, Alon U.
{ \it Network motifs in the transcriptional regulation network of
Escherichia coli} Nat Genet. 31(1):64-8 (2002)






\end{thebibliography}
\end{document}